# Impact Analysis of Allocation of Resources by Project Manager on Success of Software Projects

Gopalakrishnan Nair. T.R, Suma.V, Shashi Kumar. N.R

*Abstract*— Generation Production of successful software project is one of the prime considerations of software industry. Engineering high quality software products is further influenced by several factors such as budget, schedule, resource constraints etc. A project manager is responsible for estimation and allocation of these resources in a project. Hence, role of project manager has a vital influence on success of the project. This research comprises of an empirical study of several projects developed in a product and service based CMMI Level 5 Software Company. The investigation result shows a significant impact of aforementioned factors on the success of software and on the company. The analysis further indicates the vital role of project managers in optimizing the resource allocation towards development of software. This paper brings in impact analysis of efficiency of project manager in effectively allocating resources such as time, cost, number of developers etc. An awareness of efficiency level of project manager in optimal allocation of resources enables one to realize the desired level of quality

*Keywords*— Defect Management, Project Management, Project Manager, Software Quality, Software Engineering.

## I. INTRODUCTION

SOFTWARE has a significant role to play in almost all the domains of applications. Hence, developing high quality software is one of the key challenges of software developing centres. Quality can be viewed in various dimensions such as product level quality, process level quality, organizational level etc. Despite the existence of several quality dimensions, it is always worth to note that it is people who ultimately drive towards the production of quality in products. Hence, managing the people and resources is deemed to be one of the essential activities during software development process. Effective project management is therefore success determining strategy of software organizations [1].

However, a project manager plays a vital role in effective project management. He is accountable for the decisions upon the need of resources in terms of budget, time, technology, number of developing personnel etc for the implementation of desired application as elicited from potential stakeholders.
Hence, effective project management is more dependent on efficiency of the project manager [2].

Nevertheless, the importance of project manager in effective project management process and existence of several quality enhancing strategies, it is worth to recall that most of the project failure occurs due to inefficient project management process. This research therefore aims to analyse the impact of various resources which were allocated by project manager on the success of the project.

This paper therefore presents an empirical investigation on several projects from a product based software industry to signify the impact analysis of efficiency of project manager in effectively allocating resources. Section II of the paper briefs about the background work for this investigation. Section III describes the Research Method, Section IV presents the empirical analysis of several projects through a case study, Section V presents the Analysis and Inferences and Section VI provides the summary of this research paper.

## II. RELATED WORK

Role of project managers is one of the contributing factors which aim at developing successful project. The significant role and responsibilities of project manager is to plan, schedule and allocate resources towards development of the project. Hence, efficiency of project manager in right estimation and prediction of resources for the project has an influencing impact on the success of a project.
Authors in [3] state that the knowledge of project manager plays a critical role in the success or failure of projects. They further feel that an experienced project manager with his skill level on integration, scope, time, cost, quality, human resource, communication management and risks has an impact on the success of the project [3].

The failure of IT systems development projects has been typically be categorized as cost, time, and performance (quality) issues [4].

Authors in [1] feel that the knowledge of project manager plays a vital role in the success or failure of projects. They further express that an experienced Project Manager with his skill set comprising of integration, scope, time, cost, quality, human resource, communication management, risks and

Gopalakrishnan Nair T.R., was with Research Industry and Incubation Centre, Dayananda Sagar Institute, Bangalore, India and Aramco Endowed Chair - Technology, PMU, KSA. (e-mail: trgnair@gmail.com)
Suma. V with the Research Industry and Incubation Centre, Dayananda Sagar Institute, Bangalore, India ( e-mail: sumavdsce@gmail.com)
Shashi Kumar N.R. with the Research Industry and Incubation Centre, Dayananda Sagar Institute, Bangalore, India ( e-mail: nrshash@gmail.com)

procurement management influences the success of the project. [1].

Authors in [5] state that product quality was low with moderate team satisfaction and that with an increased team satisfaction; the quality of the product further accelerates [5].

Further, authors in [6] suggest that project team potency is influenced by project team culture and that the project success and project member satisfaction is influenced by project team potency [6].

Defect management is one of the core demands for the success of the project and it is a fact that team performance influences the effective defect management [7] [8].

Authors in [9] feel that software engineering has now taken a new perspective where project manager looks for professional management and software quality assurance methodologies during developmental activities [9].

However, author in [10] expresses that project manager has an main role in balancing and satisfying competing demands for project scope, time, cost, risk, and subsequently on quality [10].

Authors in [11] suggest that success of project depends much on project manager estimation capability of parameters like the number of defects and its presence at various phases of software development [11].

Research made by authors in [12] indicates a vital need for analytical reasoning from project managers towards effective resource allocation for defect management in order to realize successful software projects [12]

Our investigation therefore provides an empirical study of several projects to bring in awareness on the efficiency of project manager in accurate estimation and resource allocation by indicating the impact analysis of these resources on the success of a project.

## III RESEARCH WORK

Since, software has a major role in all domains of applications developing quality software is one of the needs of every industry. In order to develop quality software, role of project manager has a vital contribution in achieving the same. Hence, this research aimed to explore the impact of various resources upon the success of the project from the perspective of resource allocation decision of project managers. In order to achieve this objective, the investigation included a deep analysis of several projects developed in a leading CMMI and ISO certified product based software industry. Empirical data was collected from document management repositories. Modes of data collection included log reports, interviews, telephonic conversations, emails involving the entire project developing teams of the industry. Analysis of the data indicates the impact of various resources upon the success of the project based on the decision of their estimation and allocation by the project manager.

## IV CASE STUDY

The case study comprises of a CMMI level 5 and ISO certified service based software industry. The company functions on Business Intelligence, data warehouse, Enterprise Resource Planning, Business Process Outsourcing, Banking, Finance, Airlines and Energy Utilities.

Table 1 depicts a sampled data of fifteen projects developed since 2009 to 2012. The sampled projects are developed in .Net and Java programming languages. The table provides information about the log of estimated and actual data on project completion time, cost, defect count and total number of developers involved during developmental process.

In order to resolve the varied complexities of production, this study considers medium and large projects. Medium projects require less than 5000 hours of total software development time. Large projects consider more than 5,000 hours of total software development time. These projects are developed on Oracle database and used Java, .Net based tools in Linux Operating system environment.

## V ANALYSIS AND INFERENCES

Analysis for the empirical data is carried out on the various resources. It is worth to note that success of the project depends on quality of the product. Quality however is achieved through parameters such as cost, time, number of developers in the development team, their effort, defect management ability of the team, technology, operational environment, programming language, standard and policies etc. This part of the research focuses upon role of project manager and effectiveness of his resource allocation in the project. Hence, parameters such as cost, time, number of developers and defect count which are some of the highly quality influencing parameters are considered for the purpose of this investigation. In order to study the impact of the considered parameters towards the success of the project, this work involves incremental technique of impact analysis of each of them. Accordingly, cost parameter is compared with time upon the success level of the project. Subsequently, these two parameters are in turn analyzed with number of developers and henceforth the success of the project. Further progressing in this incremental mode, impact of number of developers is analyzed with defect count. Finally, defect count is analyzed with time, cost parameters.

Table 1 infers that as variations observed in cost parameter in terms of estimation and actual need of the cost during the project increases, the success level of the project decreases. Similar inferences may be drawn with time as parameter. However, the variations as observed in estimating the number of developers prior to the start of the project by project managers and subsequently the allocation of the number of developers during the span of developmental activities indicates unpredictable variations with success of the project. Similar inference may be drawn for the defect count with success level of the project. Hence, it is not just the number of developers or defect count variations that affect the software quality, but it is the right choice of developers and their apt effort in defect management that influences the success of the project. Further, it is also important to note that project managers are responsible in selection and allocation of resources. Our forth coming research work indicates the significance of right allocation of developers which deems the effort analysis of the same.

Figure 1 illustrates the quality of a software depends upon different parameters and our focus under study throws light on the No. of Developers, Time, Cost and defect management. Progressing further in this impact analysis of resources, our subsequent focus was towards analyzing the mutual impact of resources in the project. Figure 2 depicts the impact of variations of cost with time in the project. From the figure, it is apparent that time and cost is mutually dependent on each other.

TABLE-1: THE SAMPLED DATA OF 15 PROJECTS

| PROJECTS | NO. OF DEVELOPERS | | | DEFECTS | | | TIME | | | COST($) | | | Success Level of a Project |
|---|---|---|---|---|---|---|---|---|---|---|---|---|---|
| | E | A | %Var. | E | A | % Var. | E | A | % Var. | E | A | % Var. | % |
| P1 | 2 | 2 | 0.0 | 589 | 595 | -1.0 | 2332 | 2120 | 9.1 | 58300 | 53000 | 9.1 | 95.71 |
| P2 | 1 | 1 | 0.0 | 144 | 145 | -0.4 | 582.4 | 520 | 10.7 | 14560 | 13000 | 10.7 | 94.74 |
| P3 | 1 | 1 | 0.0 | 69 | 72 | -4.5 | 297.6 | 248 | 16.7 | 7440 | 6200 | 16.7 | 92.80 |
| P4 | 1 | 1 | 0.0 | 67 | 65 | 2.5 | 288 | 240 | 16.7 | 7200 | 6000 | 16.7 | 91.04 |
| P5 | 3 | 5 | 66.7 | 1111 | 1200 | -8.0 | 4200 | 4000 | 4.8 | 105000 | 100000 | 4.8 | 82.95 |
| P6 | 6 | 8 | 33.3 | 1944 | 1532 | 21.2 | 7700 | 7000 | 9.1 | 192500 | 175000 | 9.1 | 81.82 |
| P7 | 5 | 7 | 40.0 | 1700 | 1665 | 2.1 | 7344 | 6120 | 16.7 | 183600 | 153000 | 16.7 | 81.15 |
| P8 | 5 | 7 | 40.0 | 1789 | 1652 | 7.7 | 7599 | 6440 | 15.3 | 189980 | 161000 | 15.3 | 80.46 |
| P9 | 2 | 3 | 50.0 | 536 | 535 | 0.1 | 2314 | 1928 | 16.7 | 57840 | 48200 | 16.7 | 79.14 |
| P10 | 2 | 3 | 50.0 | 456 | 459 | -0.8 | 2017 | 1640 | 18.7 | 50430 | 41000 | 18.7 | 78.34 |
| P11 | 3 | 5 | 66.7 | 1011 | 1100 | -8.8 | 4404 | 3640 | 17.4 | 110110 | 91000 | 17.4 | 76.85 |
| P12 | 1 | 2 | 100.0 | 162 | 159 | 2.0 | 700.8 | 584 | 16.7 | 17520 | 14600 | 16.7 | 66.17 |
| P13 | 1 | 2 | 100.0 | 100 | 95 | 5.0 | 432 | 360 | 16.7 | 10800 | 9000 | 16.7 | 65.42 |
| P14 | 1 | 2 | 100.0 | 200 | 201 | -0.5 | 907.2 | 720 | 20.6 | 22680 | 18000 | 20.6 | 64.81 |
| P15 | 1 | 2 | 100.0 | 48 | 45 | 6.9 | 217.5 | 174 | 20.0 | 5437.5 | 4350 | 20.0 | 63.28 |

P: Project; E: Estimated; A: Actual; Var.: Variance

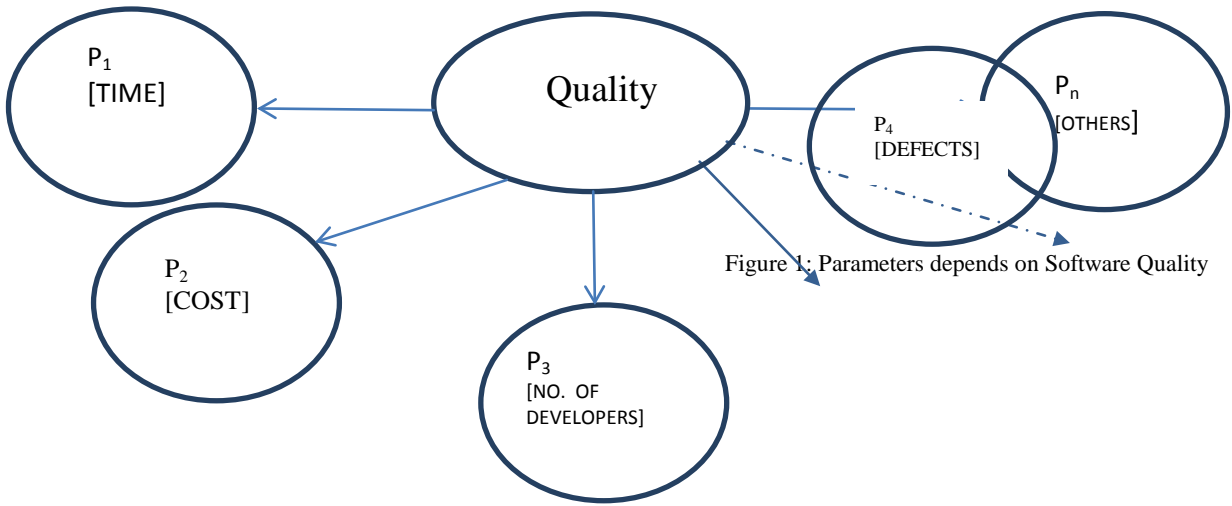

Figure 1: Parameters depends on Software Quality

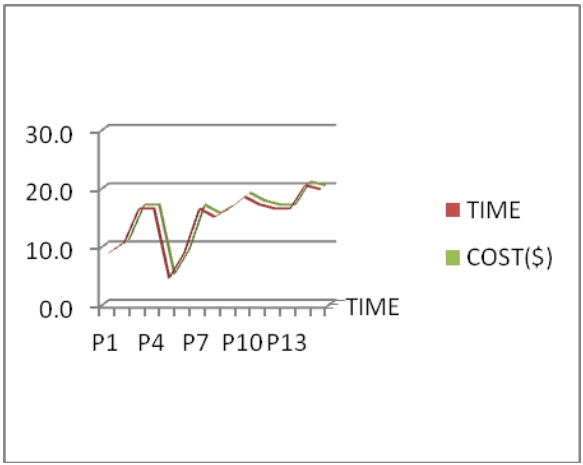

Fig. 2 Impact of variation of Time and Cost

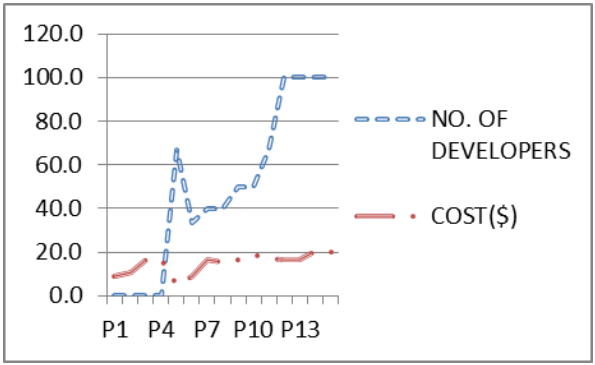

Fig. 3 Impact of variation of Developers and Cost

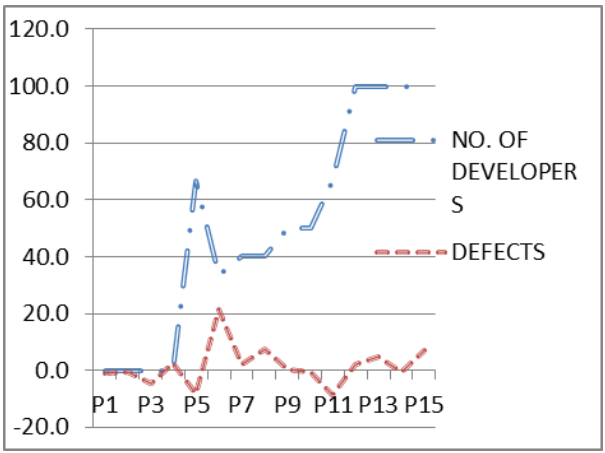

Fig. 4 Impact of Variation of No. of Developers and Defects

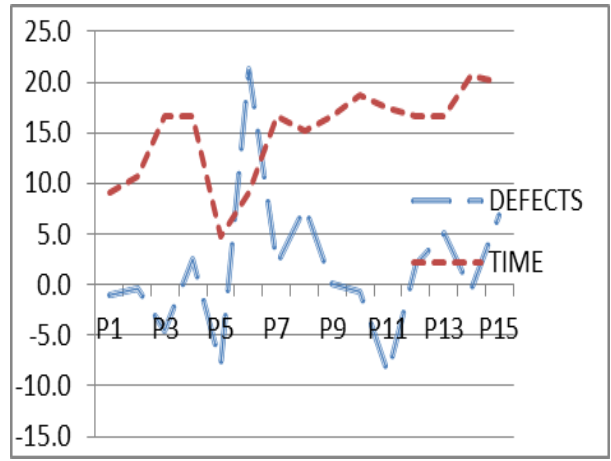

Fig. 5 Impact of Variation of Defects and Time

Figure 3 depicts the impact of variation of number of developers with cost in the projects. From the figure, it is inferred that there is hardly any noticeable impact of number. of developers on the cost.

Figure 4 depicts the Impact of Variation of number of developers with defects in the project. From the figure, it is apparent that it is not a rule of thumb that by increasing number of developers, more number of defects is captured. This has been proven from the figure where defect count has increased and has not decreased by the addition of number of developers. Our forth coming work put forth the analysis of efficiency of developers towards achieving decreased defect count as defect management is one of the influencing parameter for project success.

Figure 5 depicts the Impact of Variation of Defects and Time in the project. From the figure, it is shown that increase in number of defect count in the project proportionally total time required to complete the project. Therefore Our forth coming work put forth the analysis of efficiency of developers towards achieving decreased defect count as defect management is one of the influencing parameter for project success..

## VI    CONCLUSION

Software has become one of the widely required components of any application domain. Hence, production of high quality software is one of the core needs of any industry. Generating high quality software is dependent of various parameters which includes cost, time, number of developers, technology, and complexity of the project and so on.

Role of project manager is one of the highly modulating factors that aims towards estimation and apt allocation of resources in successfully developing projects. However, the deep investigation carried on several empirical projects developed at various software industries indicates the existence of variations between resource estimation prior to the development process and actual allocation of resources during the developmental period by the project manager. This paper therefore aims to investigate the impact of these variations of resources towards the attainment of success of the project. The parameters that are considered in this part of research include cost, time, number of developers and defect count upon the success of the project. The knowledge of impact of variations in these resources assures the organization in effectively planning, controlling and developing projects that ultimately leads towards production of high quality software which in turn guarantee completely satisfied software products.

## ACKNOWLEDGEMENT

The authors would like to acknowledge the software company involved in this study and the project managers for their invaluable help in providing necessary information for our work under the framework of the Non-Disclosure Agreement.